\documentclass[review]{elsarticle}

\usepackage{lineno,hyperref}
\usepackage{gensymb}
\modulolinenumbers[5]

\journal{Nuclear Instruments and Methods in Physics Research A}

\begin{document}

\begin{frontmatter}

\title{A facility to evaluate the focusing performance of mirrors for Cherenkov Telescopes}

\author[mymainaddress1]{Rodolfo Canestrari\corref{mycorrespondingauthor}}
\cortext[mycorrespondingauthor]{Corresponding author}
\ead{rodolfo.canestrari@brera.inaf.it}
\author[mymainaddress2]{Enrico Giro}
\author[mymainaddress1]{Giacomo Bonnoli}
\author[mymainaddress2]{Giancarlo Farisato}
\author[mymainaddress2]{Luigi Lessio}
\author[mymainaddress2]{Gabriele Rodeghiero}
\author[mymainaddress2]{Rossella Spiga}
\author[mymainaddress3]{Giorgio Toso}
\author[mymainaddress1]{Giovanni Pareschi}


\address[mymainaddress1]{INAF-Osservatorio Astronomico di Brera, via Emilio Bianchi 46, 23807 Merate, ITALY}
\address[mymainaddress2]{INAF-Osservatorio Astronomico di Padova, vicolo Osservatorio 5, 35100 Padova, ITALY}
\address[mymainaddress3]{INAF-IASF di Milano, via Bassini , Milano, ITALY}

\begin{abstract}
Cherenkov Telescopes are equipped with optical dishes of large diameter -- in general based on segmented mirrors -- with typical angular resolution of a few arc-minutes. To evaluate the mirror's quality  specific metrological systems are required that possibly take into account the environmental conditions in which typically these telescopes operate (in open air without dome protection). For this purpose a new facility for the characterization of mirrors has been developed at the labs of the Osservatorio Astronomico di Brera of the Italian National Institute of Astrophysics. The facility allows the precise measurement of the radius of curvature and the distribution of the concentred light in terms of focused and scattered components and it works in open air. In this paper we describe the facility and report some examples of its measuring capabilities.
\end{abstract}

\begin{keyword}
Cherenkov Telescopes \sep Mirrors \sep Scattering \sep Focusing
\end{keyword}

\end{frontmatter}


\section{Introduction}\label{intro}
\subsection{Scientific framework}
With the advent of the Imaging Atmospheric Cherenkov Technique (IACT) in late 1980's, ground-based observations of Very High-Energy gamma rays came into reality. Since the detection of the Crab Nebula using the IACT in 1989 by Whipple~\cite{weekes} the number of high energy gamma-ray sources has rapidly grown. Today the sources are more than 150~\cite{tevcat} and the number is increasing year by year thanks to the new generation experiments.

This first detection at TeV energies was followed by the discovery of the TeV emission from the first extragalactic source (Mrk 421), showing that acceleration processes are taking part in AGNs too~\cite{mrk421}. With the present generation experiments like H.E.S.S.~\cite{hoffmann}, VERITAS~\cite{holder} and MAGIC~\cite{ferenc} new classes of sources as well as about a dozen of unknown new ones were detected at GeV-TeV energies both galactic and extragalactic. The recent advances in $\gamma$-ray astronomy have shown that the 10 GeV -- 100 TeV energy band is crucial to investigate the physics in extreme conditions. Some interesting scientific topics are the Galactic Center, Pulsar Wind Nebulae, Pulsars and Binary Systems, Blazars, radio-galaxies, star-forming galaxies. For the interested reader, a comprehencive review on TeV Astronomy has been recently published in~\cite{tevrev}. 

Ground-based experiments using Cherenkov photons produced in air represent a cost-effective way to implement observations in this band. At present, MAGIC, H.E.S.S. and VERITAS are the state of the art of such ground-based experiments. They have collecting area, obtained by combining several mirror segments, of the order of 500-1000 $m^2$.\\
The Cherenkov Telescope Array (CTA) represents the future generation of IACT, with the goal of increasing sensitivity by a factor of 10 with respect to the present best installations and a total mirror collecting area of the array of the order of $10^4 \; m^2$. The CTA observatory is a project designed by a worldwide consortium that will make use of well demonstrated technologies of present generation Cherenkov telescopes as well as new developed solutions. CTA will be based on telescopes with different sizes installed over a large area. At its southern site e.g. 70 Small Size Telescopes (4 m primary mirror diameter), 20 Medium Size Telescopes (12 m) and 4 Large Size Telescopes (23 m) are envisaged to be implemented in order to cover a broad spectral energy range from a few tens of GeV up to more than 100 TeV~\cite{cta}. 

\subsection{Mirrors calibration for IACTs}
The mirrors for Cherenkov telescopes are in general composed by many reflecting segments to be assembled together in order to mimic the full size mirror. So far, just single reflection telescopes have been used with Davis-Cotton or parabolic layouts. In both cases the segments are in general designed with a spherical geometry and proper radius of curvature. These mirrors are also characterized by a reflectivity performance typically above 80\% (in the $300 - 550 \; nm$ energy band) but, at the same time, require angular resolution of typically a few arc-minutes, i.e. about two orders of magnitude the one of mirrors for optical astronomy. Despite the quite modest requirement in angular resolution, the distribution of the concentrated light is an important parameter in the performance of such telescopes. In fact, it has a direct impact on the measured energy and flux of gamma rays from the observed sources; and moreover in the determination of the energy threshold of the instrument~\cite{hillas}. Most of the current and future Cherenkov telescopes make use of spherical mirrors; each telescope has hundreds of segments or even thousands in the CTA case~\cite{pareschi13}. In addition, it is common to have different suppliers for the same telescope. Production and testing of such mirrors need a full characterization through appropriate facilities with suitable set-up for the testing of the prototypes and to perform the quality control during the production phase in order to cross-calibrate mirrors from different industrial pipelines. \\
Optical properties, reflecting surfaces and mechanical structure are designed aiming at obtaining the best compromise between costs and performance. Cost of the industrial production has to be sufficiently low but it has to guarantee the requirements for Cherenkov optics. To address these issues, for instance, the CTA observatory is planning to take advantage of calibration facilities. Some of those are based on the direct imaging of a light source. There are already calibration facilities based on this method available in T\"ubingen (Germany)~\cite{bernloehr}, Saclay (France)~\cite{brun} and San Antonio de los Cobres (Argentina)~\cite{medina}. Another approach which is now widely being used for mirrors, either Cherenkov or not, is based on the deflectometry method. It consists in observing the distortions of a defined pattern after its reflection by the examined surface and from them to reconstructing the surface shape. A facility based on this concept has been developed at Erlangen-N\"urnberg University~\cite{erlangen}. A variant of this method has been implemented at the Osservatorio Astronomico di Brera of the Italian National Institute of Astrophysics (INAF-OAB) to test and characterize the mirrors for the ASTRI SST-2M telescope proposed for the CTA~\cite{sironi}. A similar approach was previously used also for the characterization of mirrors for ring imaging Cherenkov counters~\cite{stutte}.

In this framework, a new optical facility has been implemented by INAF-OAB. It has been designed and developed to test spherical mirrors with long radius of curvature (several tens of meters). The facility is a system working in open-air, so that accurate evaluation of the main parameters can be achieved under different environmental condition. Moreover this facility is able to accurately investigate the scattering effect by means of an high sensitivity large format CCD camera. Several light sources with different spectral emissions are also available. 
In principle, this facility can be used either during the prototyping phase or the production phase. However, considering the high number of segments required by Cherenkov telescopes the most appropriate use of this facility is to cross-calibrate the characterization pipeline of the suppliers and to perform random checks in the production.
In this paper we present the facility and discuss its measuring capabilities.

\section{Apparatus description}
\begin{figure}[!h]
\centering
\includegraphics[keepaspectratio]{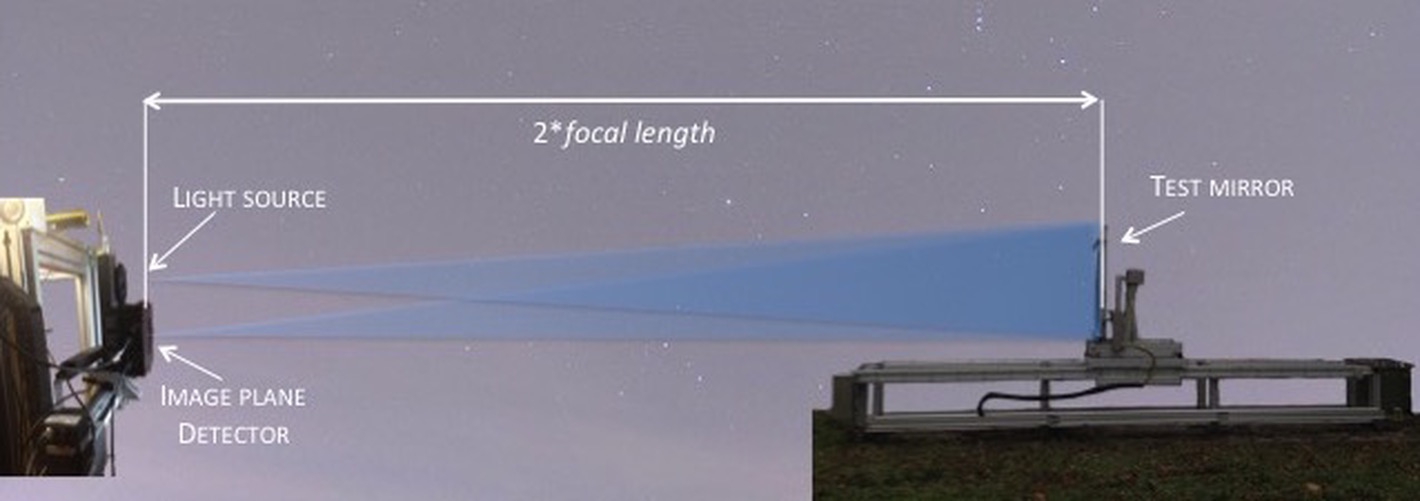}
\caption{Schematic representation of the \textit{2-f method} measurement setup.}
\label{2f}
\end{figure}

The facility measures the focused light of the mirrors using a simple optical configuration. Since mirrors have a spherical surface profile, a spherical wavefront can be used to generate the focal spot from the radius of curvature. This setup is commonly referred as \textit{2-f method}; it is sketched in Figure~\ref{2f}. To retrieve the focal length $f$ of the mirror under test the well known formula for the conjugate points can be used: 
\begin{equation}
\frac{1}{f} = \frac{1}{p} + \frac{1}{q}
\end{equation}
where $p$ is the distance of the object (e.g. a light source) from the mirror and $q$ is the distance of its image from the mirror. Assuming spherical mirrors (i.e. the typical geometry of the mirror segments used by Cherenkov telescopes), once the light source is positioned at a distance of $p = 2 \cdot f$, then the image can be seen at the same distance $q = p$, because the incoming rays hit the surface of the mirror perpendicularly and are reflected back along the incoming direction -- this distance being the radius of curvature $r = 2 \cdot f$ of the mirror under test. 

\begin{figure}[!h]
\centering
\includegraphics[keepaspectratio]{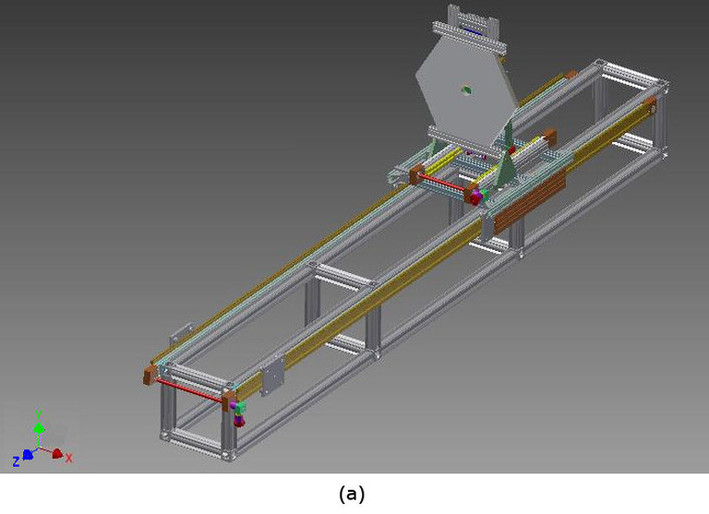}
\includegraphics[keepaspectratio]{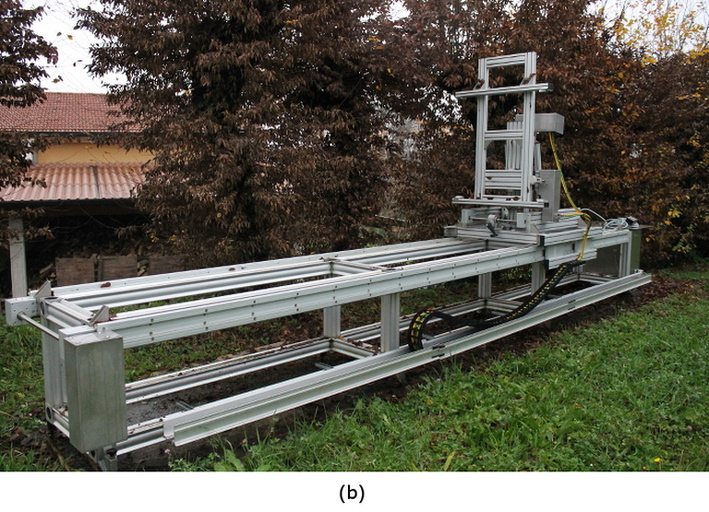}\\
\caption{The stage~\#1 of the facility: (a) 3D drawing with a mirror and (b) a photography of the system installed at the park of the Observatory.}
\label{renderingoutdoor}
\end{figure}

The above mentioned optical setup is the simplest one to check the imaging quality of the mirrors, however it requires a long baseline. The only possibility to provide a setup with a shorter length would be to produce parallel light rays which hit the surface and get focused at a distance $q = f$ from the mirror (called \textit{1-f method}). The problem with the 1-f setup is that one needs a light source emitting parallel rays which illuminate the whole mirror facet (typically larger than $1 \; m^2$), which would be much harder to realize.\\
The equipment needed to perform the \textit{2-f} test is schematically based on a light source, a detector and a room which shall be large enough to host the baseline. Our facility is indeed composed of two stages. The stage~\#1 is a mirror's support structure mounted on a long travel rail. The mirror's support and the rail are motorized in order to allow the alignment of the mirror under test. Figure~\ref{renderingoutdoor} shows a rendering of the design study performed on this part and a photo. The stage~\#2 is located into a control room where a compact bench hosting a light source and a detection unit take place. This system is motorized, thus enabling the possibility to scan the focusing plane. A control-command unit (i.e a desktop computer), an electrical cabinet and storage space complete the apparatus.\\
The facility is installed at the Merate (Lecco, Italy) site of INAF-OAB. It is based on a long baseline to fit mirrors with radii of curvature ranging from 30 meters up to 36 meters. This choice was driven by the fact that most of the current and future Cherenkov telescopes (e.g. H.E.S.S., MAGIC and CTA) make use of mirrors with similar characteristics. Moreover, the stage~\#1 is installed outdoor, thus giving the possibility to study also the mirror performance for different thermal conditions, i.e. mimicking the real operative configuration of the mirrors mounted on a real Cherenkov telescope.

As previously stated, this method is widely used for the characterization of mirrors for Cherenkov telescopes. However the facility presented in this paper has few peculiar characteristics that, combined together, makes it unique and innovative. These features are:
\begin{itemize}
\item the entire system has been designed to be user-friendly. To this regard, the manipulator hosting the mirror and the support of the detector are fully robotized. They can be easily automated to run long-time acquisitions without the on-site intervention of the operators. The automation reduces the time needed to fully characterize a mirror to no more than 15 minutes;
\item the stage~\#1 is installed in open air. Cherenkov telescopes typically work between $-10$\degree C - $+30$\degree C and 5-90\% relative humidity. Indeed, their mirrors are not influenced from temperature variations of few degrees as those experienced during a typical nighttime period. While the long time variations such as the seasonal ones (of the order of 30-40\degree C) could in principle change the radius of curvature up to few percent; hence the focusing performance of the mirrors. The seasonal variation effects are what the facility can investigate;
\item the direct imaging on a large format CCD camera mounted on a 2-axis motorized stage. This configuration is a high sensitivity setup that allows to catch diffused photons on a large area and perform a correct evaluation of the Encircled Energy function of the mirror. This kind of study is of great importance for the evaluation of the large deviations from the ideal focal position due to the scattering from the micro-rough profile of the mirror.
\end{itemize}
In the following subsections we report a detailed technical description of the two stages.

\subsection{The stage \#1, outdoor}
The outdoor stage has three main subsystems: a rail, a mirror's support and an electrical cabinet. It has been conceived and designed at the INAF-OAB. The engineering, realization and installation activities were performed by the Officina Opto-Meccanica Insubrica and Automation One companies~\cite{oomi}. 

\begin{figure}[!h]
\centering
\includegraphics[keepaspectratio]{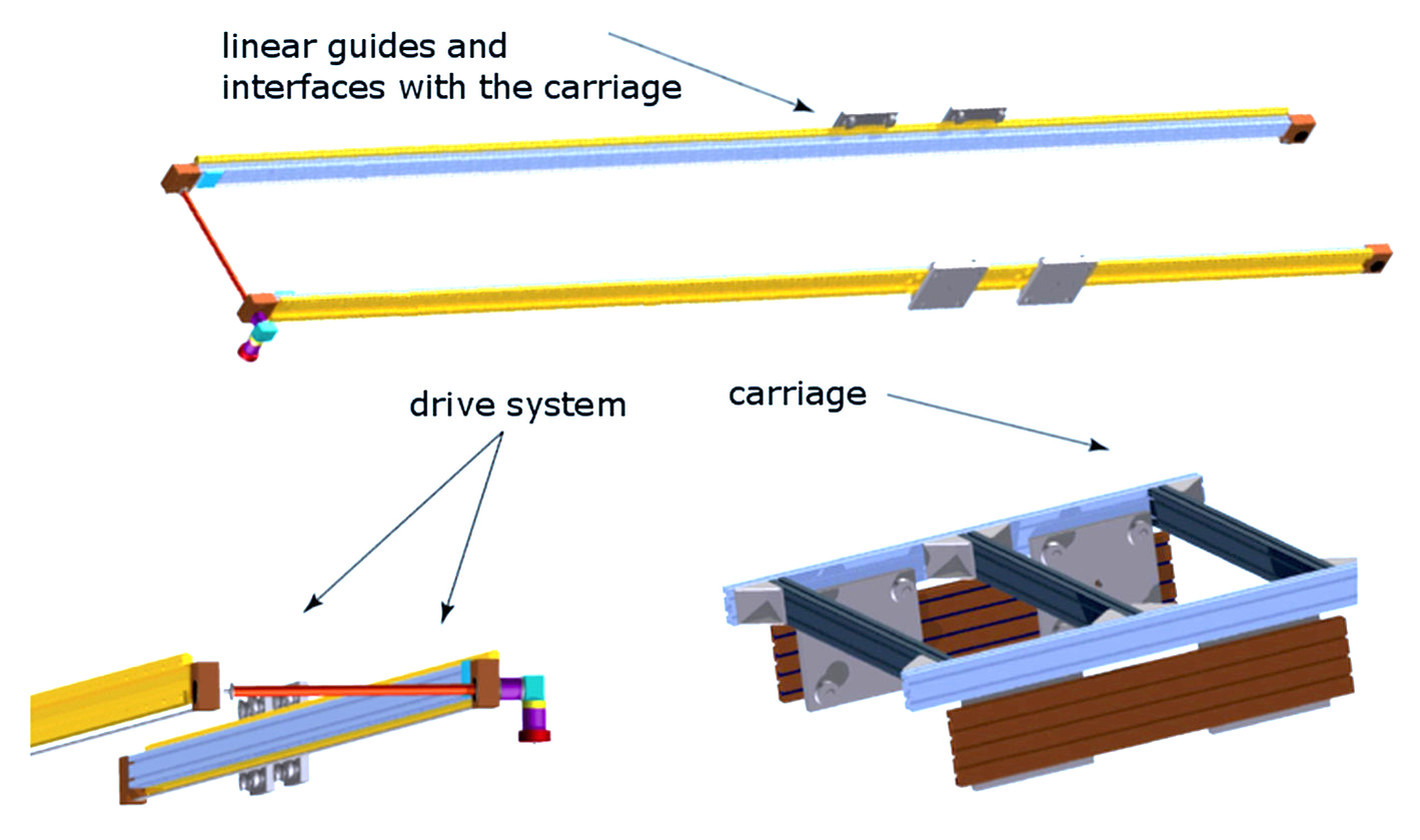}\\
\caption{Detailed view of the rail subsystem. It is composed of two 6~m long linear guides, a carriage and the drive system.}
\label{rail-cad}
\end{figure}

\paragraph{The rail subsystem}
It is composed of a couple of 6 m long stainless steel linear guides. The drive system uses one brushless motor with IP code 65 and can work in open environment. Its rear shaft is equipped with an absolute rotary encoder with EnDat interface. The shaft of the motor brings the motion to both the linear guides and it is then distributed to the carriage through toothed belts. This solution is able to ensure a positioning of the carriage well below 1~mm on the full travel range of the rail, because the position loop is closed through the reading of the encoder. The nominal position of the carriage with respect to the indoor stage of the facility is recorded by an external laser distance meter, it is suitable for outdoor measurements over large distances. It guarantees a knowledge of the optical baseline within a few mm. The rail subsystem is mounted over an optical bench made of aluminum profiles. Figure~\ref{rail-cad} details the rail subsystem.\\

\begin{figure}[!h]
\centering
\includegraphics[keepaspectratio]{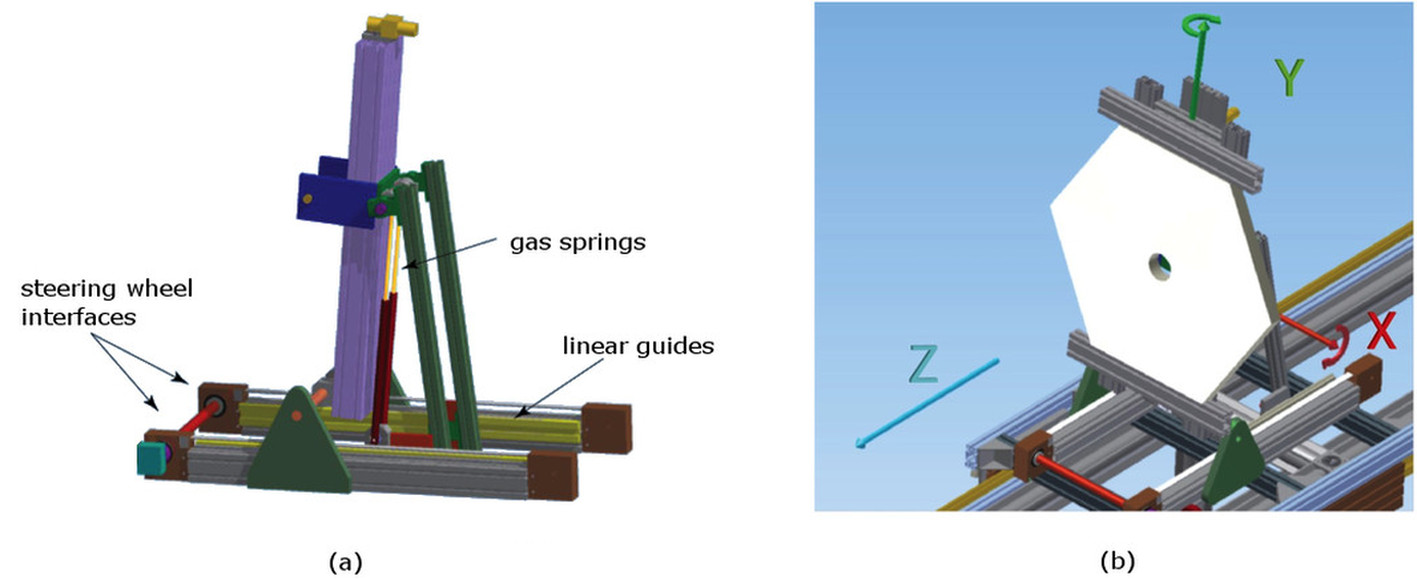}\\
\caption{Detailed view of the mirror's support subsystem. Panel (a) shows the support itself; panel (b) shows the reference axes for the three motorized motions of the outdoor stage.}
\label{mirrorsupport-cad}
\end{figure}

\paragraph{The mirror support}
The mirror support is shown in Figure~\ref{mirrorsupport-cad}. It is installed over the carriage and is designed to ease the mounting and dismounting operations of the mirror under test as well as to facilitate the alignment of the mirror itself over the long optical baseline of the facility. It can be divided into two parts. One part can be horizontally reclined to execute the loading and unloading of the mirror. When this part is standing vertical, it can be blocked to prevent undesired movements. 

The holding for mirrors is obtained by means of an adjustable system of aluminum beam profiles and soft clamps. This part can be moved in such a way the mirror tilts with respect to two axes. The drive system is based on linear actuators with re-circulating ball screws. It has wide angular ranges of 5\degree \ along the x axis and 10\degree \ along y, with a resolution better than 12 arcsec. The mirror's support can be loaded up to 45~kg, different mirror tile's shapes (e.g. squares, hexagons, rounds) up to 1.5~meters in diameter can be managed. 

\paragraph{The outdoor electrical cabinet}
The electrical cabinet is made of a stainless steel water tight box for external applications. It is equipped with a thermoregulation system composed by heaters, coolers and dryers to keep the electronics within its working conditions. This solution ensures the functionality of the facility within a wide range of environmental conditions.

In addition to the thermoregulation system, the cabinet hosts the drivers to pilot the three motors of the motion system, an ethernet switch and a gateway to handle the input/output digital signals. The cabinet receives the power from the main grid of the Observatory through a dedicated 400V line. The power is handled by the system to provide 220V and 24V lines that are distributed to all the devices of the outdoor stage whether they are resident in the cabinet or not. A safety stop red push button is available for emergency handling. The cabinet is equipped with a proper interface to connect a keypad to send motion commands to the system.

\subsection{The stage \#2, indoor}
The indoor stage is based on three main subsystems: a light source, a photon detection unit and an electrical cabinet. 

\begin{figure}[!h]
\centering
\includegraphics[keepaspectratio]{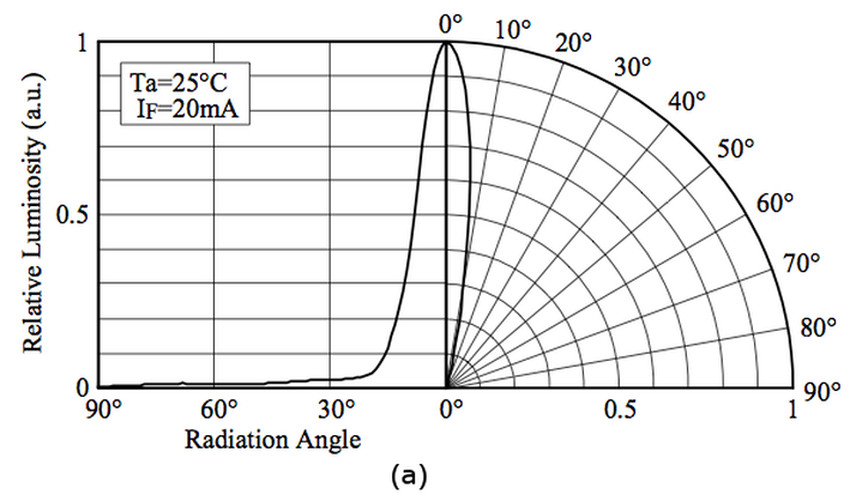}
\includegraphics[keepaspectratio]{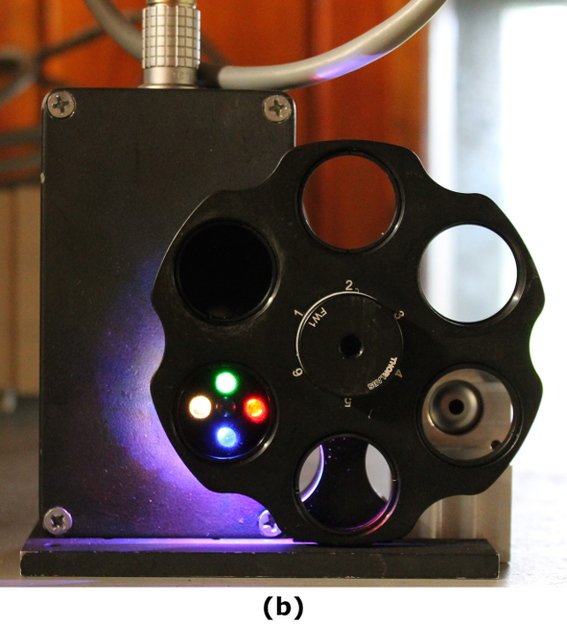}\\
\caption{(a) The light emission diagram (light cone and related intensity) of the LEDs used and (b) the light source unit assembled with the RGB LED and the laser beam switched off.}
\label{led}
\end{figure}

\paragraph{The light source}
The light source is a compact device able to generate a spherical wavefront of light. Five ultra-bright LED sources are disposed in a pattern: an RGB LED is surrounded by a red (626~nm), a green (525~nm), a blu (470~nm) and a warm white LEDs. Any combination of LEDs can be switched ON and OFF, as needed for the measurement.
With respect to laser sources, the choice of LEDs has been made as a compromise for their cheapness and safety of use versus the quality of the wavefront generated (quality in terms of light intensity, spatial distribution and emission angle). Concerning the quality of wavefront, the half cone emission diagram of the LEDs used is shown in Figure~\ref{led}(a). Considering the typical angular size of the mirrors under test (i.e. 3\degree or less), this diagram shows that the non-uniformity of the light wavefront at the mirror pupil is never exceeding 2\%. A filter wheel with logarithmic neutral filters permits to dim the light intensity and illuminate the mirror with a suitable light flux in order not to saturate the detector (see Figure~\ref{led}(b)).\\
The source is equipped also with a very low power laser beam for alignment purposes.

\paragraph{The detection unit} 
This unit is composed of a CCD camera and a long-range laser distance meter for outdoor applications. The laser meter gives the absolute measurement of the distance between the detector plane and the mirror. The device is a DISTO$^{TM}$ D8 model~\cite{disto}, with a declared precision better than $\pm 5$~mm up to 36~m.

\begin{figure}[!h]
\centering
\includegraphics[keepaspectratio]{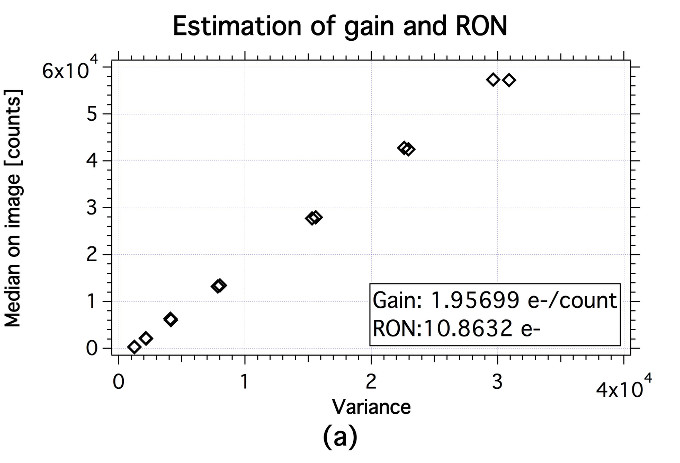}
\includegraphics[keepaspectratio]{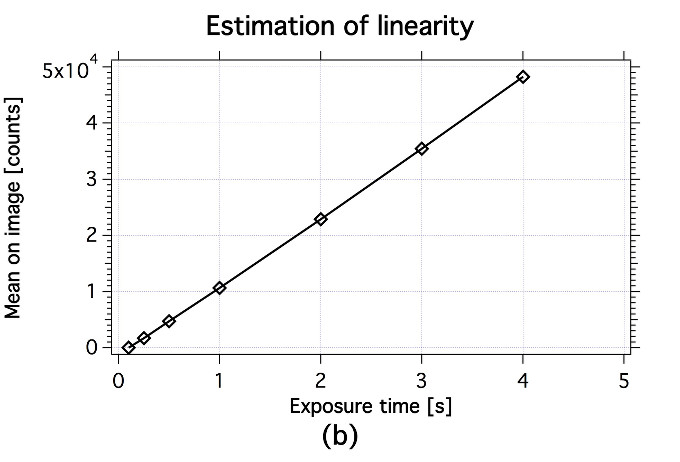}\\
\includegraphics[keepaspectratio]{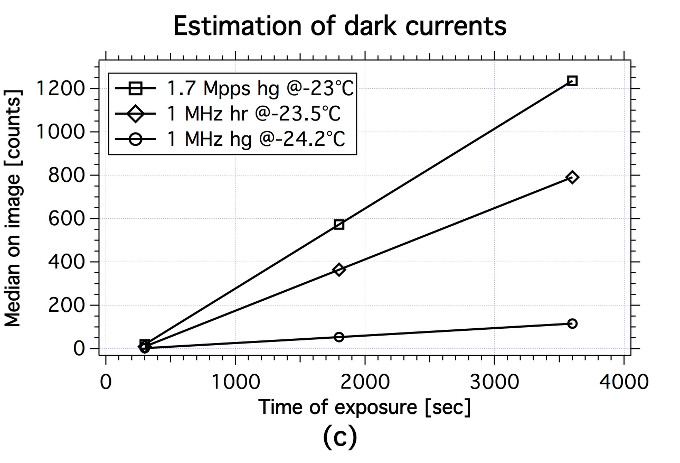}
\includegraphics[keepaspectratio]{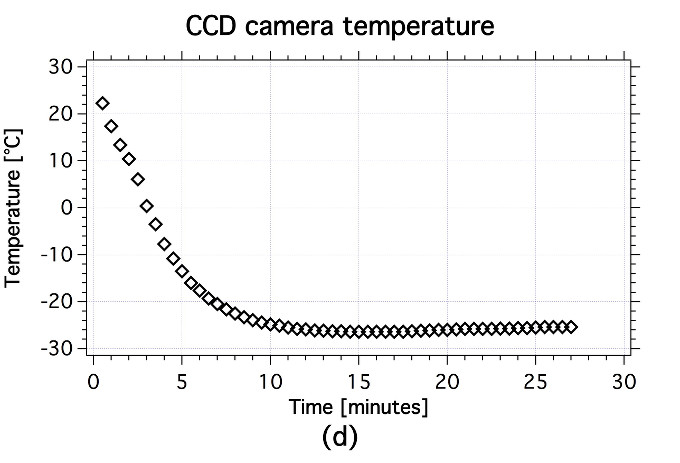}\\
\caption{(a) Gain and RON estimation, (b) linearity (c) dark currents and (d) typical evolution of the temperature during cooling time for the entire range.}
\label{CCDcalib}
\end{figure}

\begin{figure}[!h]
\centering
\includegraphics[keepaspectratio, width= 7cm]{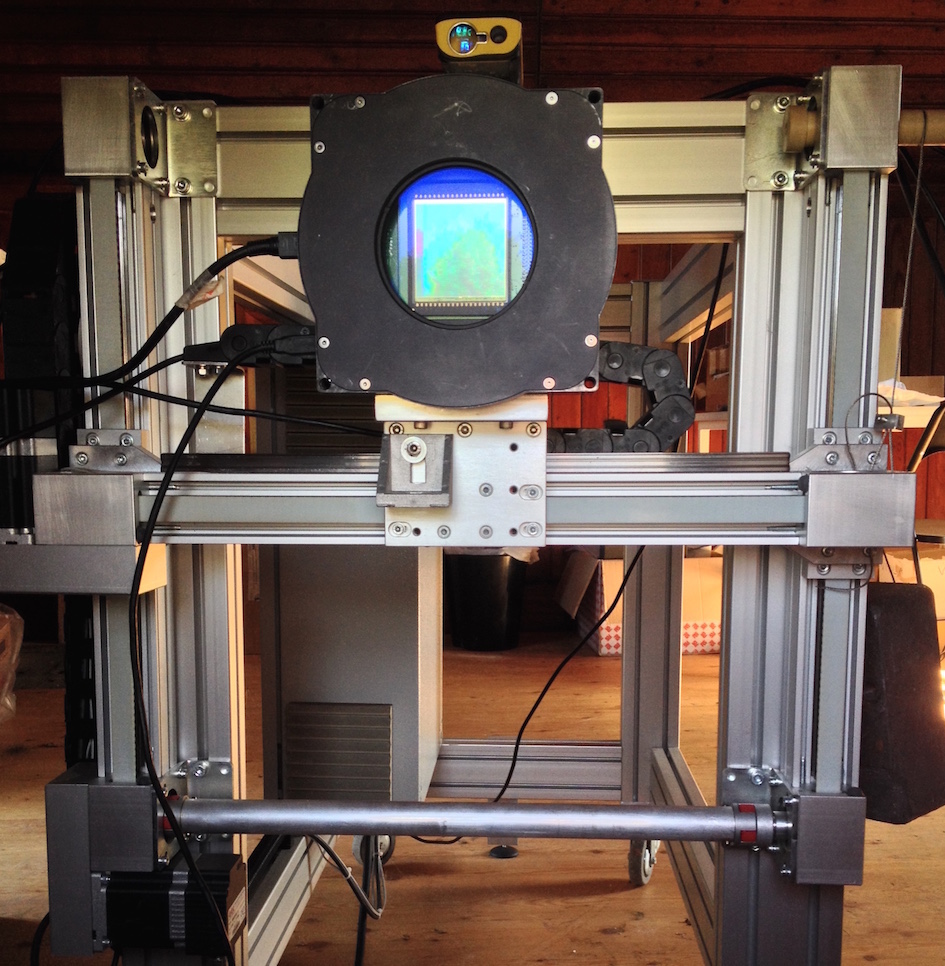}
\caption{Picture of the detection unit: the main parts are the 2-axis stage, the CCD camera with open shutter and the laser meter.}
\label{detunit}
\end{figure}

The CCD camera is used to detect the light reflected back from the mirror under test. The model is PL4301E from the ProLine series~\cite{fli} characterized by low noise, high sensitivity, high resolution and deep cooling. The sensor mounted is a Truesense KAF-4301 from ON Semiconductor Inc. producer~\cite{CCD}. It is a large format CCD with $2048 \times 2048$ pixels 24~$\mu$m side for a total diagonal of 70.7~mm. The camera is equipped with a 90~mm shutter to avoid any vignetting on the detector. A filter wheel can be mounted on top either to dim the incoming light or to select a particular wavelength, in case of need. The PL4301E has a thermoelectric cooling system capable to cool down the detector temperature to 50\degree C below the ambient one (see Figure~\ref{CCDcalib}(d)).\\
The CCD camera has undergone a careful characterization in terms of gain (named also conversion factor $e^-/count$), Read-Out Noise (RON), linearity, dark current and Charge Transfer Efficiency (CTE). The gain has been evaluated acquiring a series of images with increasing exposure times followed by another series of images with decreasing times. For each image the variance is computed and plotted against the median counts of the image itself. The gain corresponds to the angular coefficient of the best fit line, while the RON is obtained by multiplying the gain and the mean value of the bias frame. Each image used is given by the mean of two subsequent acquisitions A and B, both corrected for dark and bias signals. This procedure guarantees to check both shot- and long- term variations of the camera.\\
Linearity and dark current are evaluated by varying the exposure time. Respectively, by acquiring a number of bright and dark acquisitions at increased exposure times. Then the median counts of the acquired images is computed and plotted against the exposure time~\cite{abbott}.\\
The CTE has been derived by the cosmic rays impacts detected after a 1800 seconds dark exposure. Cosmic rays impact the detector as stochastic events with casual angles and energy but they can be used to diagnose the CTE of the detector as suggested by A. Riess et al.~\cite{cte}.\\
All these parameters depend on the download speed (i.e. the frame readout frequency). We report in Figure~\ref{CCDcalib} the results for the 1~MHz high gain setup that is typically used. In table~\ref{CCDcalibtab} the full set of calibration results are reported.\\
The detection unit is completed by a 2-axis stage to move around the CCD camera along the detection plane. The scan covers an area of $280 \times 290 \; mm^2$. The motion has a resolution of 0.1~mm. Figure~\ref{detunit} shows the system assembled.

\begin{table}[!h]
\caption{Analytical results of the CCD calibration for different download speeds.}
\centering
\begin{tabular}{lcccc}
\textbf{Download speed} 	& \textbf{Gain} 	& \textbf{RON} 	& \textbf{Dark currents} 	& \textbf{CTE}	\\
					& [$e^-/counts$] & [$e^-$]		& [$e^-/pixel/s$]		& [\%]		\\
\hline
1 MHz high gain		& 1.95699		& 10.8632		& $<0.03$				& 99.999878	\\
1 MHz high range		& 12.6399		& 37.4377		& $<0.2$				& 99.999731	\\
1.7 Mpps high gain		& 1.73786		& 13.0477		& $<0.3$				& 99.999769	\\
\hline
\end{tabular}
\label{CCDcalibtab}
\end{table}%

\paragraph{The indoor electrical cabinet}
This cabinet routes all the I/O commands and communication signals between the computer and the outdoor stage. Its main components are the ethernet switch and the gateway. It has an independent 220V power line with respect to the outdoor cabinet. The main power is handled by the system to provide proper voltages to all its internal devices. A safety red push stop button is available for emergency handling.  Also this cabinet has the interface to connect a keypad to send motion commands to the system.

\subsection{The software interface}
To run the complete facility four independent programs are available: one for the outdoor stage, two for the detection unit and one for the light source. However, not all these computer programs are mandatory to acquire measurements, since the choice depends on the kind of measure the user is interested in.\\ 
Concerning the detection unit, one application is devoted to the motion of the axes. It can be programmed to make follow a specific path. The CCD detector has its own software that permits a variety of acquisitions and settings (e.g. the dark and flat field frames, binning mode, gain etc.), it commands also the filter wheel. Both are commercially available software programs that come with the hardware.\\
The light source can be commanded to switch on the different LEDs with which it is equipped.

With the outdoor stage control program commands and setting instructions can be transmitted. It shows three panels, one for each axis of the motion. The user can set the maximum speed of the motion and the position to be reached, either in absolute or relative. The user can also set the motion in jog mode, this configuration being particularly useful during the alignment phase.\\
Each axis can be independently set with respect to the other ones; more axes can be moved at the same time.

\section{Measurements and calibrations with the facility}
In this section we discuss some typical measurements and calibrations that can be pursued with the facility. All the results are based on the photometric analysis of the data retrieved and on the information of the distance read by the laser distance meter.

\subsection{Evaluation of \textsf{r80} and best focus position}\label{r80}
In analogy with an optical telescope, the angular resolution quality of a mirror for Cherenkov telescope is evaluated from its Point Spread Function (PSF). However, the parameter in general used for the Cherenkov case is the \textsf{r80}, i.e. the radius that contains the 80\% of the focused light. This parameter is tipically preferred with respect to the more commonly used (in optical astronomy) Full Width Half Maximum (FWHM). Indeed the PSF of the mirrors for Cherenkov telescopes can hardly be reducible to a Gaussian distribution since the shape's errors introduced from the low cost manufacturing processes adopted are dominant with respect to the intrinsic aberration of the theoretical design. The micro-roughness can also play an important role. Moreover, since Cherenkov observations often deal with very faint signals, the \textsf{r80} turns to be a better estimator to qualify the mirrors and hence the amount of concentrated light.

A standard measurement carried out with the facility is the acquisition of a number of images at various distances from the mirror. After the mounting of the mirror on its support and the alignment of the optical axis with the light source and detector unit (\textit{z} axis in Figure~\ref{mirrorsupport-cad}), the procedure foresees the rough localization of the best focus position. All the measurements are then taken with discrete steps around this position. The discrete steps are measured by means of the encoder mounted on the shaft of the \textit{z} axis motor. The value of the origin in terms of distance from the mirror is retrieved by averaging a number of reads with the laser distance meter. Particular care is taken in settling this point in order to avoid systematic errors.\\
Each image is then treated as an astronomical image and analyzed following standard aperture photometry procedures (i.e. use of dark and bias frames, evaluation of the background, etc.) and using standard software routines (e.g. Daophot photometry library~\cite{astrolib}, SAOImage DS9~\cite{ds9}, etc.).\\
\begin{figure}[!h]
\centering
\includegraphics[keepaspectratio]{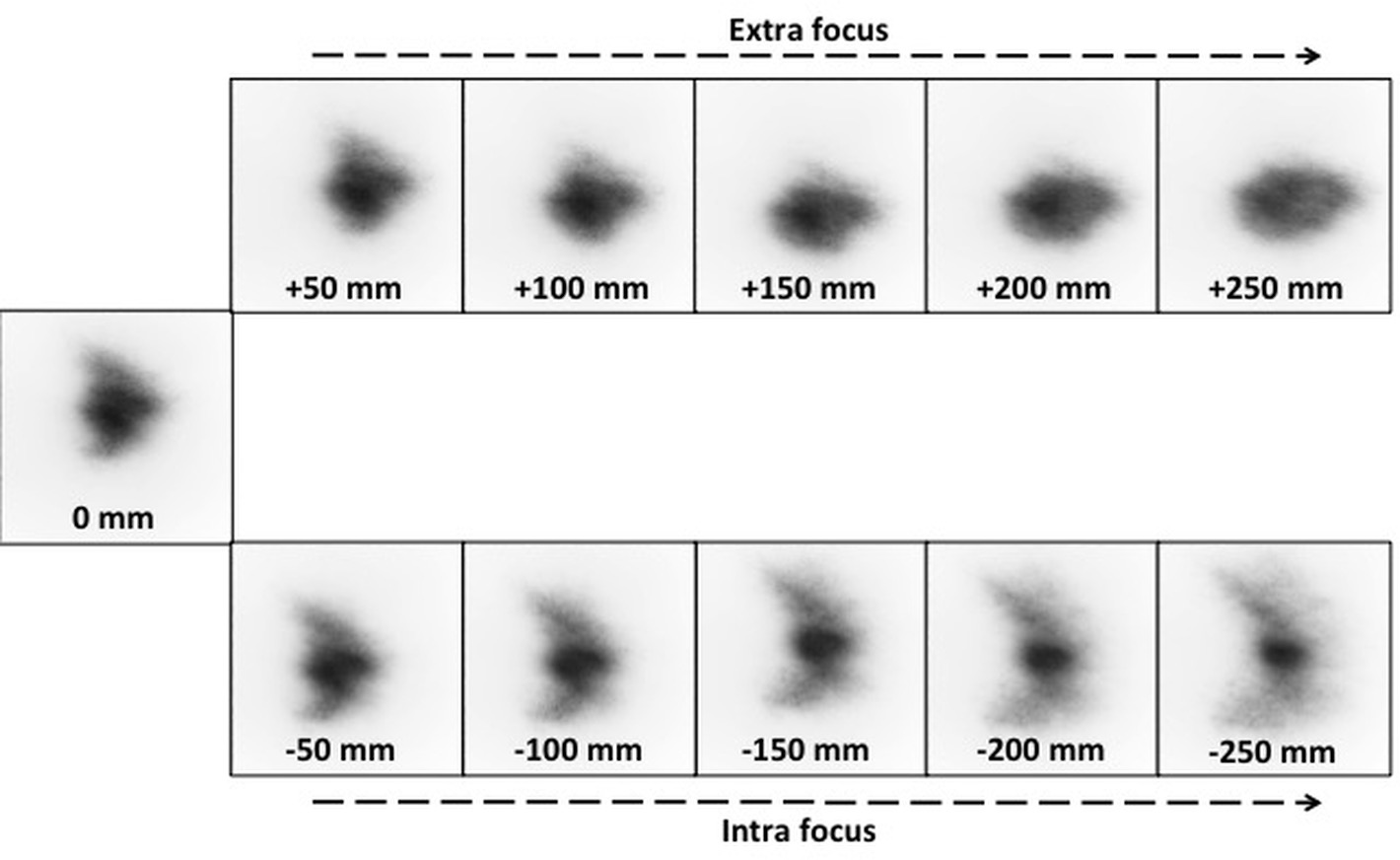}
\caption{PSFs generated by the mirror along its optical axis for the full measuring length.}
\label{PSFbestfocus}
\end{figure}
\begin{figure}[!h]
\centering
\includegraphics[keepaspectratio]{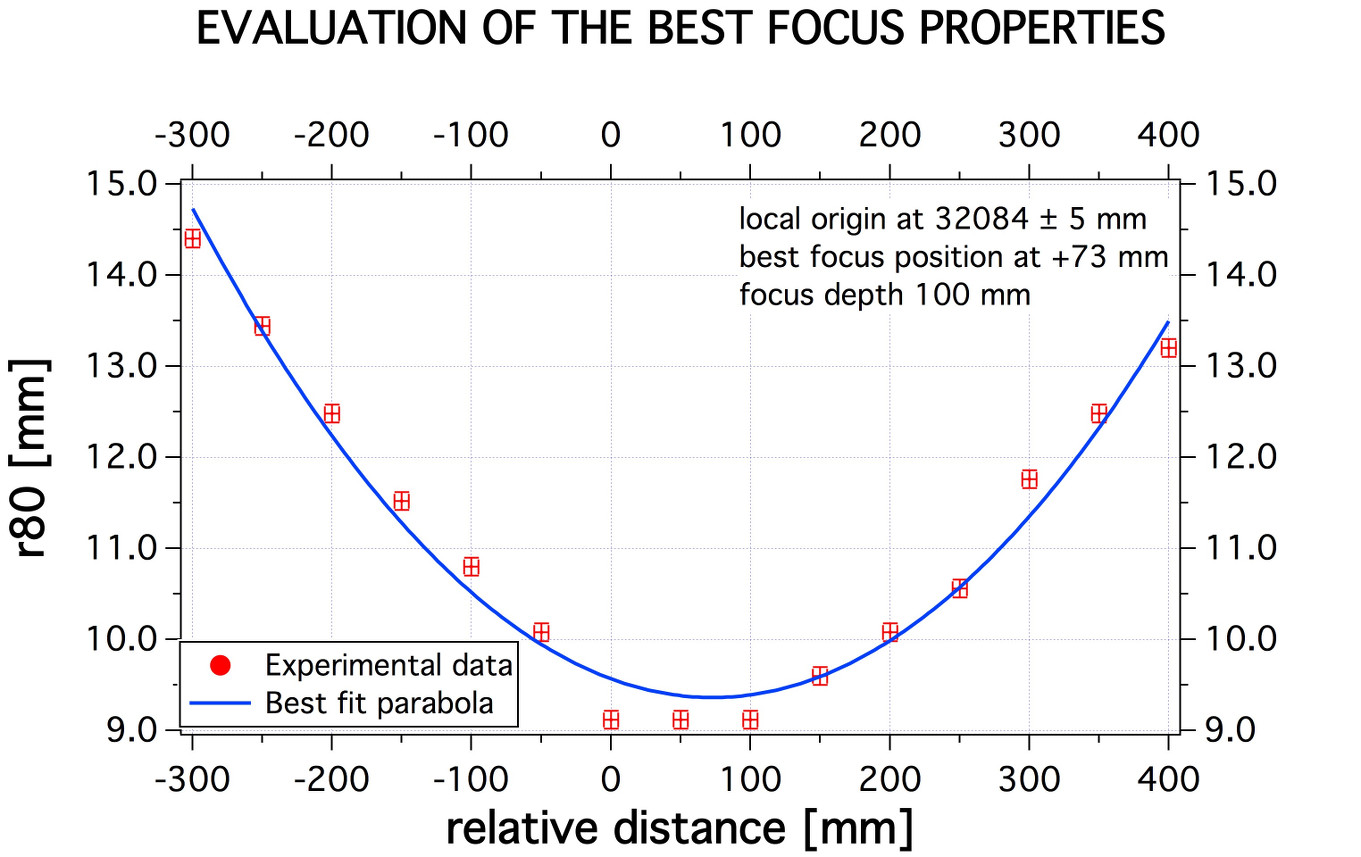}\\
\caption{Evaluation of the best focus and depth of focus values for a mirror by means of the \textsf{r80} property. With this facility it is possible to define the best focus position within few per thousand.}
\label{focalparabola}
\end{figure}
An example of the results is shown in Figure~\ref{PSFbestfocus} and Figure~\ref{focalparabola}, respectively: the serie of PSF images taken at the various distances from the origin and the results of the images analysis. In particular, from the plot in Figure~\ref{focalparabola} it is possible to estimate two geometrical parameters of the mirror under test: the best focus position (being also the radius of the best fitting sphere) and the focal depth. The first parameter is evaluated from the vertex of the parabola that best fits the experimental data, while the latter is due to the sensitivity in estimating the \textsf{r80} and its relative uncertainty from the experimental PSFs data. The errors associated to the \textsf{r80} and the relative distance are evaluated in two different ways. For the \textsf{r80} we use the Poissonian noise associated both with the PSF photometry and the background evaluation. The two values are quadratically summed, even if typically the second one is the dominant contribution. For the relative distance the sensitivity of the DISTO$^{TM}$ D8 is taken.\\ 
In the example shown in Figure~\ref{focalparabola} we obtained the best focus position at +73~mm from the local origin with a focal depth of 100~mm (over a radius of curvature of $32084 \pm 5$~mm). For the \textsf{r80} values we have computed an error of $\pm 0.1 \; mm$.

\subsection{Evaluation of the astigmatism}
\begin{figure}[!h]
\centering
\includegraphics[keepaspectratio]{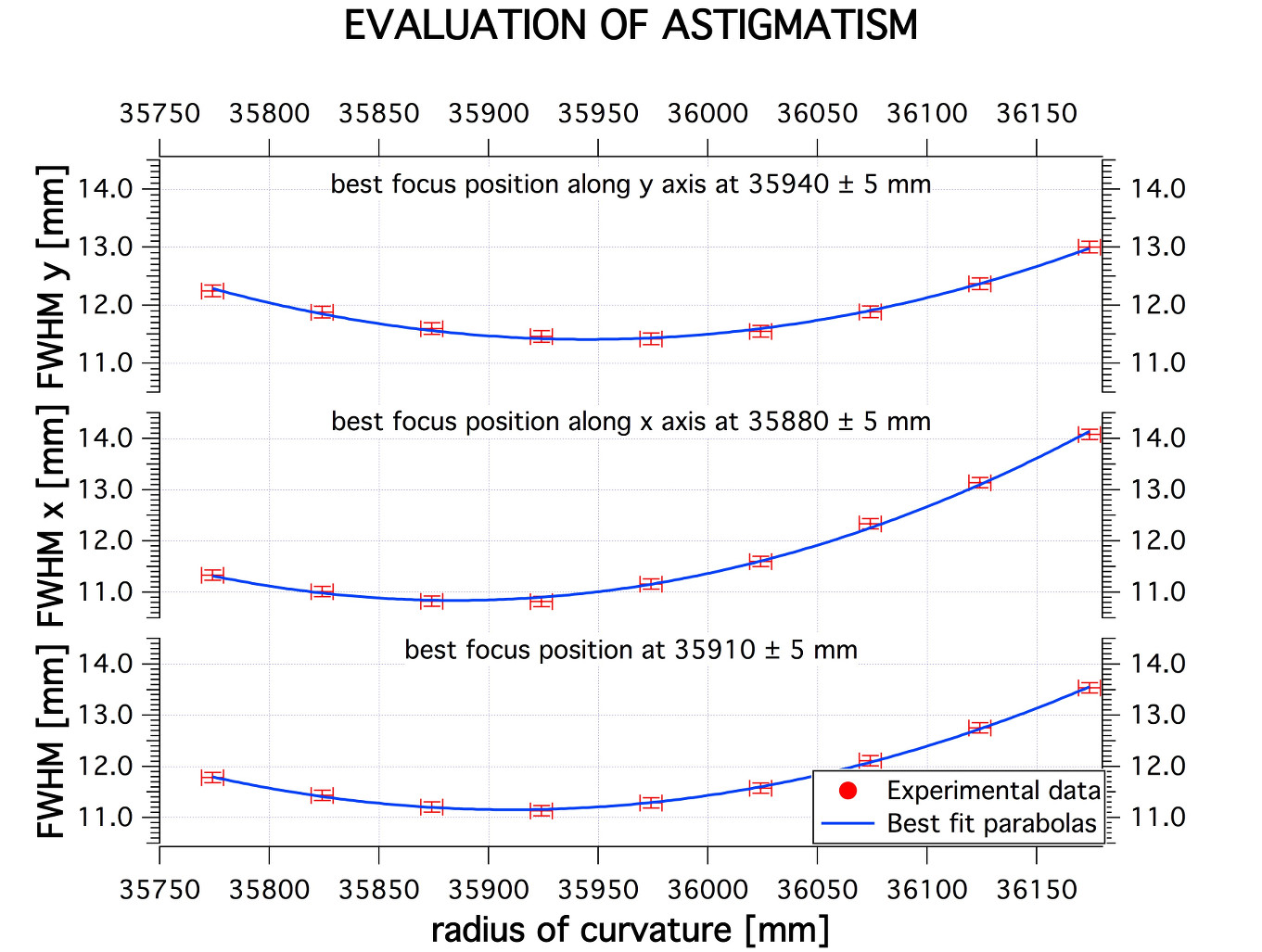}
\caption{Evaluation of the astigmatism aberration as function of the radius of curvature by measure of the FWHM along two orthogonal axis. Total FWHM is also shown.}
\label{astigatism}
\end{figure}

More detailed investigation on the shape's errors of the mirrors can be undertaken using the FWHM as estimator. By evaluating the contributions over the two orthogonal axes lying on the focus plane (\textit{x} and \textit{y} axes in Figure~\ref{mirrorsupport-cad}) it is possible to disentangle the astigmatism aberration of the mirror.\\ 
The procedure to acquire the measures is the same as that described in Section~\ref{r80}, the analysis is also based on standard aperture photometry, but FWHM is taken as reference instead of \textsf{r80}.\\
In Figure~\ref{astigatism} we present, from bottom to top, the plots of the total FWHM, the FWHM along the \textit{x} axis and the FWHM along the \textit{y} axis as functions of the focal distance (the radius of curvature). Experimental data and best fit parabolas are shown. It is possible to appreciate a difference in the best focus positions independently achieved on the \textit{x} and \textit{y} axes of 60~mm, over a radius of curvature of $35910 \pm 5$~mm.

\subsection{Scattering evaluation}\label{scat}
The diffuse scattering is, in general, due to irregularities of the surface of the mirror at microscopic level that induce coherent large angular deviations from the specular direction, thus generating a broad diffused light component surrounding the core of the PSF. If those irregularities have a specific spatial pattern, the scattering can generate structured tails in the PSF. The more pronounced the irregularities are, the more diffused the light is, thus covering a wide area on the focus plane and reducing the amount of light falling into the telescope camera detector. The method to detect the scattering is very important to understand the behavior of the mirror in terms of angular resolution. 

To cover a wider area around the PSF we therefore raster scan the focal plane. For each position an image is acquired that is later stitched together with the others to generate a wide single image of the focus plane.

Different approaches have been suggested to evaluate the integral value of the specular plus scattered components~\cite{razmik} that require an ad hoc setup. The one proposed here makes use of the same equipment as for \textsf{r80}. Moreover, aperture's photometry directly on the CCD is profitably exploited to avoid using any objective or imaging screen that would imply a transfer function.

\begin{figure}[!h]
\centering
\includegraphics[keepaspectratio]{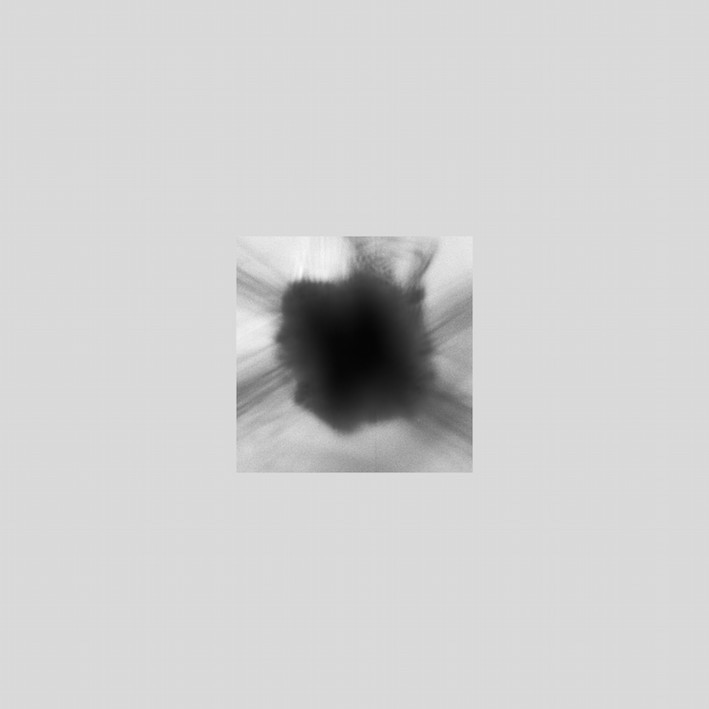}
\includegraphics[keepaspectratio]{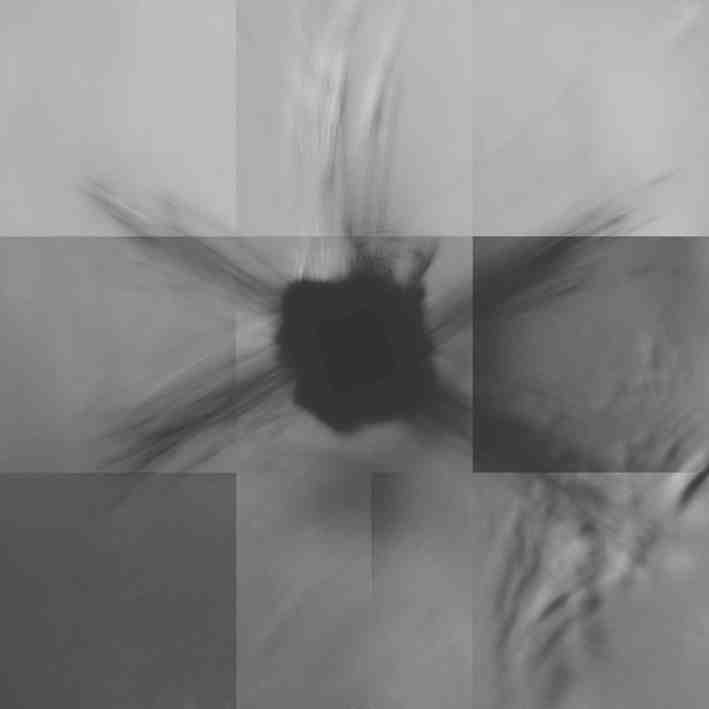}\\
\caption{(a) PSF acquired by a single frame of the CCD located at the center of the detector plane and (b) PSF acquired by means of a composition of 9 frames of the CCD obtained with a raster scan of the detector plane. The image's contrast is intentionally exaggerated to highlight the scattering component.}
\label{scattering}
\end{figure}

\begin{figure}[!h]
\centering
\includegraphics[keepaspectratio]{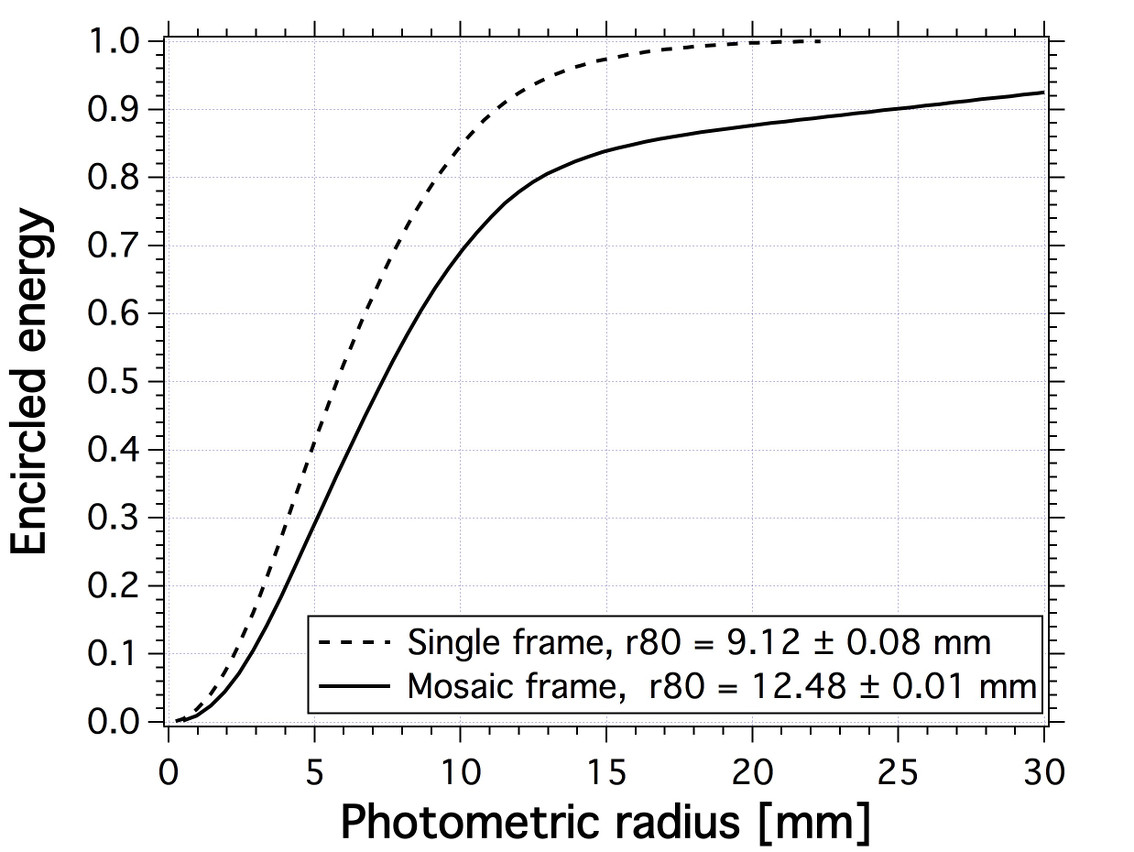}\\
\caption{Evaluation of the \textsf{r80} value from a 1-frame PSF image and a 9-frame PSF image of the same mirror.}
\label{r80mosaic}
\end{figure}

As an example of the application, we show the image of the PSF acquired by a single frame at the center of the focal plane (Figure~\ref{scattering} left panel) and by a raster scan (Figure~\ref{scattering} right panel). In the second case, the contrast has been intentionally stretched in order to saturate the bulk of the PSF (in this way the tails due to surface imperfections are clearly visible). While these tails do not influence by a reasonable amount the estimation of the best focus position, they have some effects on the total amount of concentrated light. To give the reader a quantitative value we chose a mirror with a pronounced contribution from the scattering and we compared the \textsf{r80} obtained from the two images. From the single frame we obtained \textsf{r80}$_{1frame} = 9.1 \pm 0.1$ mm while from the raster scan we got \textsf{r80}$_{9frames} = 12.5 \pm 0.01$ mm. The plot of the encircled energy is shown in Figure~\ref{r80mosaic}.

\section{Future developments}
In-focus total reflectivity is among the most important parameters for understanding the performance of a mirror for Cherenkov telescopes, indeed one of the most difficult to assess. While the local surface reflectivity is commonly measured sampling the mirror's surface with spectrophotometer devices, their detector's acceptance angle is in general wide enough to collect also an important fraction of the scattered component, mixing it to the specular reflection one. The mirror's surface shape quality is obtained through the use of facilities based on the \textit{2-f method}, as described in this paper. The capability to combine together the afore-mentioned information by means of a single measurement (now wavelength dependent) will allow us to get a more reliable evaluation of the expected PSF of the entire Cherenkov telescope and to estimate the background component due to the optical surface errors.\\ 
Such a measurement is possible thanks to the facility presented in this paper as soon as the scattering evaluation method presented in Section~\ref{scat} is coupled with a reliable way to measure the light flux of the source in use. This can be done for instance by using a calibrated photodiode and a semi-reflective folding mirror. A detailed study is ongoing and some preliminary tests have been already carried out.

Activities to improve the software programs integration are also ongoing. This will give an easier and faster measuring experience.

\section{Conclusions}
An open-air user-friendly facility for the characterization of mirrors for Cherenkov telescopes with long radius of curvature is presented. It is devoted to the precise determination of the radius of curvature and the measurement of the on-focus light distribution generated by the mirror under test. The latter in terms of focused and scattered components, normalized to the total incoming light at the detector.\\
The facility has a flexible light source able to provide wavefronts at different wavelengths. This capability combined to the large field of view of the camera and the possibility to perform raster scans, makes the facility ideal to pursue calibrations of Cherenkov mirrors with direct CCD imaging, with a correct evaluation of the  Encircled Energy function.\\ 
A detailed technical description covering its electro-mechanical, electrical, optical and software components has been presented. Some typical measurements made possible through the facility have been discussed together with the forthcoming possibility to implement the on-focus total reflectivity evaluation.\\ 
The radius of curvature and the on-focus light distribution measurements can be correlated to the ambient and/or mirror temperature thus opening the possibility to experimentally assess the thermal behavior of the mirror.

The facility is run by the INAF personnel of the Observatories of Brera and Padova but the access is open to the entire scientific community who may feel the need of these types of measurements, such as that of the CTA observatory or others present and future projects.

\section*{Acknowledgments}
This work was supported in part by the ASTRI ``Flagship Project" financed by the Italian Ministry of Education, University, and Research (MIUR) and led by the Italian National Institute of Astrophysics (INAF). We also acknowledge MIUR for the support through PRIN 2009 and TeChe.it 2014 special grants. The authors also thank Officina Opto-Meccanica Insubrica and Automation One companies for their valuable support.

\end{document}